\documentclass[journal,twocolumn,10pt]{IEEEtran}
\usepackage[T1]{fontenc}% optional T1 font encoding
\usepackage{cite}
\usepackage[cmex10]{amsmath}
\usepackage{amsthm}
\usepackage{amssymb}
\usepackage{mathtools}

\usepackage{algorithmic}
\usepackage{textcomp}
\usepackage[usenames, dvipsnames]{color}
\usepackage[linesnumbered,ruled]{algorithm2e}
\usepackage{subcaption}
\usepackage{graphicx}  % remove 'demo' option for your real document
\usepackage{url}
\usepackage{xcolor}
\usepackage[usestackEOL]{stackengine}

\DeclareMathOperator*{\argmin}{arg\;min}
\DeclareMathOperator*{\argmax}{arg\;max}

\ifCLASSINFOpdf
\else
\fi

\begin{document}
\title{SVM-based Channel Estimation and Data Detection for One-Bit Massive MIMO Systems}

\author{Ly~V.~Nguyen,  A.~Lee~Swindlehurst, and Duy~H.~N.~Nguyen
	\thanks{Ly V. Nguyen is with the Computational Science Research Center, San Diego State University, San Diego, CA, USA 92182 (e-mail: vnguyen6@sdsu.edu).}
	\thanks{A. Lee Swindlehurst is with the Department of Electrical Engineering and
		Computer Science, Henry Samueli School of Engineering, University of
		California, Irvine, CA, USA 92697 (e-mail: swindle@uci.edu).}
	\thanks{Duy H. N. Nguyen is with the Department of Electrical and Computer Engineering, San Diego State University, San Diego, CA, USA 92182 (e-mail: duy.nguyen@sdsu.edu).}}

\maketitle

\begin{abstract}
The use of low-resolution Analog-to-Digital Converters (ADCs) is a practical solution for reducing cost and power consumption for massive Multiple-Input-Multiple-Output (MIMO) systems. However, the severe nonlinearity of low-resolution ADCs causes significant distortions in the received signals and makes the channel estimation and data detection tasks much more challenging. In this paper, we show how \textit{Support Vector Machine} (\textit{SVM}), a well-known supervised-learning technique in machine learning, can be exploited to provide efficient and robust channel estimation and data detection in massive MIMO systems with one-bit ADCs. First, the problem of channel estimation for uncorrelated channels is formulated as a conventional SVM problem. The objective function of this SVM problem is then modified for estimating spatially correlated channels. Next, a two-stage detection algorithm is proposed where SVM is further exploited in the first stage. The performance of the proposed data detection method is very close to that of Maximum-Likelihood (ML) data detection when the channel is perfectly known. We also propose an SVM-based joint Channel Estimation and Data Detection (CE-DD) method, which makes use of both the to-be-decoded data vectors and the pilot data vectors to improve the estimation and detection performance. Finally, an extension of the proposed methods to OFDM systems with frequency-selective fading channels is presented. Simulation results show that the proposed methods are efficient and robust, and also outperform existing ones.
\end{abstract}


% \fi
\IEEEpeerreviewmaketitle

\section{Introduction}
The development of wireless communications systems has been moving toward the use of more and more antennas at the transceivers. Massive Multiple-Input-Multiple-Output (MIMO) technology is a result of this development and is now considered to be one of the disruptive technologies of 5G networks~\cite{Boccardi2014Five,andrews2014will}. The first and foremost benefit of massive MIMO is the significant increase in the spatial degrees of freedom obtained by combining tens to hundreds of antennas at the base station. This benefit of spatial degrees of freedom helps improve the throughput and energy efficiency by several orders of magnitude over conventional MIMO systems~\cite{Hoydis2013massive,ngo2013energy}. However, the use of many antennas at the base station also poses a number of problems. More specifically, a massive MIMO system requires many Radio-Frequency (RF) chains and Analog-to-Digital Converters (ADCs) to support a massive number of antennas. This causes significant increases in hardware complexity, system cost, and power consumption.

Recently, low-resolution ADCs have attracted significant research interest and are considered to be a promising solution for the aforementioned problems. This is due to the simple structure and low power consumption of low-resolution ADCs. As reported in~\cite{walden1999analog}, the power consumption of an ADC is exponentially proportional to its resolution. Hence, using low-resolution ADCs can significantly reduce the power consumption of the system. The simplest architecture involving 1-bit ADCs requires only one comparator and does not require an Automatic Gain Control (AGC). Thus, 1-bit ADCs are an attractive potential solution for the problems of hardware complexity, system cost, and power consumption.

One major drawback of $1$-bit ADCs is the induced distortion, since only the \textit{sign} of the real and imaginary parts of the received signals is retained. This severe nonlinearity makes the channel estimation and data detection tasks much more challenging. MIMO channel estimation with $1$-bit ADCs has been studied intensively in a number of papers with different scenarios, e.g.,~\cite{choi2016near,Risi2014Massive,li2017channel,Shilpa2019Massive,Zhichao2019Oversampling,Zhichao2019Channel,Kang2017Power,Liu2020Angular,Kim2018Dominant,Kim2018Channel,kim2019channel2,Srinivas2019Itervative,Mezghani2018Blind,kim2019channel,Mo2018Channel,Rodriguez2016Channel,Rusu2015Adaptive,Rao2019Channel}. Maximum-Likelihood (ML) and Least-Squares (LS) channel estimators were proposed in~\cite{choi2016near} and~\cite{Risi2014Massive}, respectively. The Bussgang decomposition is exploited in~\cite{li2017channel} to form a Bussgang-based Minimum Mean-Squared Error (BMMSE) channel estimator. The work in~\cite{Shilpa2019Massive} proposes a BMMSE channel estimator for massive MIMO systems with $1$-bit spatial sigma-delta ADCs in a spatially oversampled array or for sectorized users. Channel estimation with temporally oversampled $1$-bit ADCs is studied in~\cite{Zhichao2019Oversampling} and~\cite{Zhichao2019Channel}. The use of spatial and temporal oversampling $1$-bit ADCs was shown to help improve the channel estimation accuracy but requires more resources and computations due to the oversampling process. A channel estimation method based on Support Vector Machine (SVM) with $1$-bit ADCs, referred to as soft-SVM, was presented in~\cite{Kang2017Power}. Angular-domain estimation for MIMO channels with $1$-bit ADCs was studied in~\cite{Liu2020Angular,Kim2018Dominant,Kim2018Channel}. Other scenarios involving spatially/temporally correlated channels or multi-cell processing with pilot contamination were investigated in~\cite{kim2019channel2} and~\cite{Srinivas2019Itervative}, respectively. For sparse millimeter-wave MIMO channels, the ML and maximum a posteriori (MAP) channel estimation problems were studied in~\cite{Mezghani2018Blind} and~\cite{kim2019channel}, respectively. Taking into account the sparsity of such channels, the $1$-bit ADC channel estimation problem has been formulated as a compressed sensing problem in~\cite{Mo2018Channel,Rodriguez2016Channel,Rusu2015Adaptive}. Several performance bounds
on the channel estimation of mmWave massive MIMO channels with $1$-bit ADCs were reported in~\cite{Rao2019Channel}. In this paper, we focus on a more general channel model which is not assumed to be sparse without any oversampling.

Data detection in massive MIMO systems with $1$-bit ADCs has also been studied intensively in the literature, e.g.,~\cite{choi2016near},~\cite{Jeon2018One,wen2016bayes,nguyen2019linear,nguyen2019supervised,Jeon2018supervised,Kim2019SemiSupervised,jeon2019robust,Song2019CRC-Aided,Cho2019OneBitSCSO,Shao2018Iterative}. The one-bit ML detection problem is formulated in~\cite{choi2016near}. For large-scale systems where ML detection is impractical, the authors in~\cite{choi2016near} proposed a so-called near-ML (nML) data detection method. The ML and nML methods are however non-robust at high Signal-to-Noise Ratios (SNRs) when Channel State Information (CSI) is imperfectly known. A One-bit Sphere Decoding (OSD) technique was proposed in~\cite{Jeon2018One}. However, the OSD technique requires a preprocessing stage whose computational complexity for each channel realization is exponentially proportional to both the number of receive and transmit antennas. The exponential computational complexity of OSD makes it difficult to implement in large scale MIMO systems. Generalized Approximate Message Passing (GAMP) and Bayes inference are exploited in~\cite{wen2016bayes} but the proposed method is sophisticated and expensive to implement. A number of linear receivers for massive MIMO systems with $1$-bit ADCs are presented in~\cite{nguyen2019linear} and several learning-based methods are also proposed in~\cite{Jeon2018supervised,nguyen2019supervised,Kim2019SemiSupervised,jeon2019robust}. The linear receivers in~\cite{nguyen2019linear} are easy to implement but their performance is often limited by an error floor. The learning-based methods in~\cite{Jeon2018supervised,nguyen2019supervised,Kim2019SemiSupervised} are blind detection methods for which CSI is not required, but they are restricted to MIMO systems with a small number of transmit antennas and only low-dimensional constellations. Several other data detection approaches were proposed in~\cite{jeon2019robust,Song2019CRC-Aided,Cho2019OneBitSCSO,Shao2018Iterative}, but they are only applicable in systems where either a Cyclic Redundancy Check (CRC)~\cite{jeon2019robust,Song2019CRC-Aided,Cho2019OneBitSCSO} or an error correcting code such as Low-Density Parity-Check (LDPC) code~\cite{Shao2018Iterative} is available.

In this paper, we propose channel estimation and data detection methods which are efficient, robust, and applicable to large-scale systems without the need for CRC or error correcting codes. Our work is based on \textit{Support Vector Machine} (\textit{SVM}), a well-known supervised-learning technique in machine learning~\cite{bishop2006pattern}. Since SVM problems can be solved by very efficient algorithms~\cite{joachims2006training,keerthi2005modified,bottou2007support}, the proposed methods can be implemented in an efficient manner. Our earlier work reported in \cite{Ly-ICC2020} examined SVM-based channel estimation and data detection methods for $1$-bit MIMO systems with independent and identically distributed (i.i.d.) channels. This paper extends the study in \cite{Ly-ICC2020} and presents the following contributions:
\begin{itemize}
	\item An SVM-based channel estimation method for uncorrelated channels is first proposed by formulating the $1$-bit ADC channel estimation problem as an SVM problem. Unlike the soft-SVM method in~\cite{Kang2017Power}, the proposed method exploits the original idea of SVM by maximizing the margin achieved by the linear discriminator. For spatially correlated channels, we develop a new channel estimation problem by revising the conventional SVM objective function. Numerical results show that the high-SNR Normalized Mean-Squared Error (NMSE) floor of the proposed channel estimation methods is lower than that of the BMMSE method proposed in~\cite{li2017channel}, which outperforms other existing methods.
	\item We then propose a two-stage SVM-based data detection method, where the first stage is also formulated as an SVM problem. A second stage is then employed to refine the solution from the first stage. Simulation results show that the performance of the proposed method is very close to that of the ML detection method if perfect CSI is available. With imperfect CSI, the proposed data detection method is shown to be robust and to also outperform existing methods. We also provide an explanation for the non-robustness of ML detection at high SNRs with imperfect CSI.
	\item Next, an SVM-based joint Channel Estimation and Data Detection (CE-DD) method is proposed. The key idea is to exploit both the to-be-decoded data vectors and pilot data vectors to improve the channel estimation accuracy and thus improve the data detection performance.
	\item Finally, an extension of the proposed methods to OFDM systems with frequency-selective fading channels is derived. Numerical results show that the proposed SVM-based methods significantly outperform existing ones. For example, the high-SNR NMSE floor of the proposed SVM-based channel estimation method is about~$3$-dB lower that of the BMMSE method.
\end{itemize}

The rest of this paper is organized as follows: Section~\ref{sec_system_model} introduces the assumed system model. In Section~\ref{sec_proposed_methods}, we present linear SVM for binary classification and the proposed methods for flat-fading channels. Extension of the proposed methods to OFDM systems with frequency-selective fading is presented in Section~\ref{sec_extension_to_wideband}. Numerical results are given in Section~\ref{sec_Numerical_Results}, and Section~\ref{sec_conclusion} concludes the paper.

\textit{Notation}: Upper-case and lower-case boldface letters denote matrices and column vectors, respectively. $\mathbb{E}[\cdot]$ represents expectation. Depending on the context, the operator $|\cdot|$ is used to denote the absolute value of a number, or the cardinality of a set. $\|\cdot\|$ denotes the $\ell_2$-norm of a vector. The transpose and conjugate transpose are denoted by $[\cdot]^T$ and $[\cdot]^H$, respectively. The notation $\Re\{\cdot\}$ and $\Im\{\cdot\}$ respectively denotes the real and imaginary parts of the complex argument. If $\Re\{\cdot\}$ and $\Im\{\cdot\}$ are applied to a matrix or vector, they are applied separately to every element of that matrix or vector. $\mathbb{R}$ and $\mathbb{C}$ denote the set of real and complex numbers, respectively, and $j$ is the unit imaginary number satisfying $j^2=-1$. $\mathcal{N}(\cdot,\cdot)$ and $\mathcal{CN}(\cdot,\cdot)$ represent the real and the complex normal distributions respectively, where the first argument is the mean and the second argument is the variance or the covariance matrix. The operator $\operatorname{blockdiag}(\mathbf{A}_1,\ldots,\mathbf{A}_n)$ represents a block diagonal matrix, whose main-diagonal blocks are $\mathbf{A}_1,\ldots,\mathbf{A}_n$.
\section{System Model}
\label{sec_system_model}
\begin{figure}[t!]
	\centering
	\includegraphics[width=\linewidth]{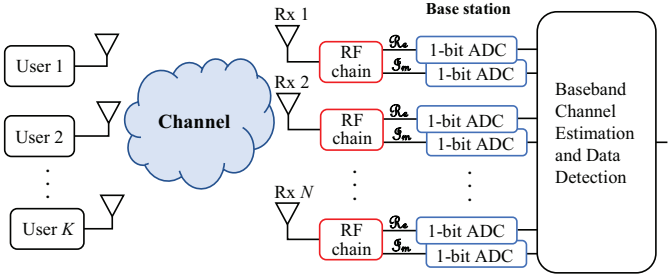}
	\caption{Block diagram of a massive MIMO system with $K$ single-antenna users and an $N$-antenna base station equipped with $2N$ $1$-bit ADCs.}
	\label{fig_system_model}
\end{figure}
We consider a massive MIMO system as illustrated in Fig.~\ref{fig_system_model} with $K$ single-antenna users and an $N$-antenna base station, where it is assumed that $N \geq K$. Let $\bar{\mathbf{x}} = [\bar{x}_1, \bar{x}_2, \ldots, \bar{x}_K]^T \in \mathbb{C}^K$ denote the transmitted signal vector, where $\bar{x}_k$ is the signal transmitted from the $k^{\text{th}}$ user under the power constraint $\mathbb{E}[|\bar{x}_k|^2]=1$, $k \in \mathcal{K} = \{1, 2, \ldots, K\}$. Let $\bar{\mathbf{H}} \in \mathbb{C}^{N\times K}$ denote the channel, which for the moment is assumed to be block flat fading. Let $\bar{\mathbf{r}} = [\bar{r}_1, \bar{r}_2, \ldots, \bar{r}_N]^T \in \mathbb{C}^N$ be the unquantized received signal vector at the base station, which is given as 
\begin{equation}
\bar{\mathbf{r}} = \bar{\mathbf{H}}\bar{\mathbf{x}}+\bar{\mathbf{z}},
\end{equation}
where $\bar{\mathbf{z}} = [\bar{z}_1, \bar{z}_2, \ldots,\bar{z}_N]^T \in \mathbb{C}^{N}$ is a noise vector whose elements are assumed to be i.i.d. as $\bar{z}_i \sim \mathcal{CN}(0,N_0)$, and $N_0$ is the noise power. Each analog received signal $\bar{r}_i$ is then quantized by a pair of $1$-bit ADCs. Hence, we have the received signal
\begin{equation}
\bar{\mathbf{y}} = \operatorname{sign}(\bar{\mathbf{r}}) = \operatorname{sign}\left(\Re\{\bar{\mathbf{r}}\}\right) + j\operatorname{sign}\left(\Im\{\bar{\mathbf{r}}\}\right)
\end{equation}
where $\operatorname{sign}(\cdot)$ represents the $1$-bit ADC with $\operatorname{sign}(a) = +1$ if $a \geq 0$ and $\operatorname{sign}(a) = -1$ if $a < 0$. The operator $\operatorname{sign}(\cdot)$ of a matrix or vector is applied separately to every element of that matrix or vector. The SNR is defined as $\rho = 1/N_0$.

\section{Proposed SVM-based Channel Estimation and Data Detection with $1$-bit ADCs}
\label{sec_proposed_methods}
\subsection{Linear SVM for Binary Classification}
Consider a binary classification problem with a training data set of $P$ data pairs $\mathcal{D} = \left\{(\mathbf{x}_q,y_q)\right\}_{q=1,\ldots,P}$ where $\mathbf{x}_q$ is a training data point and $y_q \in \{\pm1\}$ is an associated class label. Note that $\{\mathbf{x}_q\}$ here are vectors of real elements. The data set $\mathcal{D}$ is said to be linearly separable if and only if there exists a linear function $f(\mathbf{x}) = \mathbf{w}^T\mathbf{x} + b$ such that $\forall q\in \{1,2,\ldots, P\}$, $f(\mathbf{x}_q) > 0$ if $y_q = +1$ and $f(\mathbf{x}_q) < 0$ if $y_q = -1$.
Here, $\mathbf{w}$ and $b$ are referred to as the weight vector and the bias, respectively. In other words, the hyperplane $f(\mathbf{x}) = \mathbf{w}^T\mathbf{x} + b = 0$ divides the space into two regions where $f(\mathbf{x})=0$ acts as the \textit{decision boundary}. The margin of the hyperplane $f(\mathbf{x})=0$ with respect to $\mathcal{D}$ is defined as
\begin{equation}
m_{\mathcal{D}}(f) = \frac{2}{\|\mathbf{w}\|}.
\end{equation}

The SVM technique seeks to find $\mathbf{w}$ and $b$ such that the margin $m_{\mathcal{D}}(f)$ is maximized. The optimization problem can be expressed as~\cite{bishop2006pattern}
\begin{equation}
\begin{aligned}
& \underset{\{\mathbf{w},b\}}{\operatorname{minimize}}
& & \frac{1}{2}\|\mathbf{w}\|^2\\
& \operatorname{subject\ to}
& & y_q(\mathbf{w}^T\mathbf{x}_q+b)\geq 1, \quad q = 1, 2, \ldots, P.
\end{aligned}
\label{eq_SVM_problem}
\end{equation}
In case the training data set $\mathcal{D}$ is not linearly separable, a generalized optimization problem is considered as follows:
\begin{equation}
\begin{aligned}
& \underset{\{\mathbf{w},b,\xi_q\}}{\operatorname{minimize}}
& & \frac{1}{2}\|\mathbf{w}\|^2+C\sum_{q=1}^{P}\xi_q\\
& \operatorname{subject\ to}
& & y_q(\mathbf{w}^T\mathbf{x}_q+b)\geq 1 - \xi_q,\\
& & & \xi_q \geq 0, \quad q = 1, 2, \ldots, P.
\end{aligned}
\label{eq_generalized_SVM_problem}
\end{equation}
Here, $\{\xi_q\}$ are slack variables and $C>0$ is a parameter that ``controls the trade-off between the slack variable penalty and the margin"~\cite{bishop2006pattern}. The optimization problems~\eqref{eq_SVM_problem} and~\eqref{eq_generalized_SVM_problem} can be solved by very efficient algorithms~\cite{joachims2006training,keerthi2005modified}. For example, if the weight vector is sparse, the complexity of the algorithm in~\cite{joachims2006training} scales linearly in both the number of features (size of the weight vector $\mathbf{w}$) and the number of training samples $|\mathcal{D}|$. For arbitrary weight vectors, the complexity of the algorithm in~\cite{keerthi2005modified} scales linearly in the number of training samples and quadratically in the number of features for the worst case. A good review of efficient methods for solving~\eqref{eq_SVM_problem} and~\eqref{eq_generalized_SVM_problem} can also be found in~\cite{bottou2007support}.

\subsection{Proposed SVM-based Channel Estimation}
\subsubsection{Uncorrelated Channels}
First, we consider uncorrelated channels where the channel elements are assumed to be i.i.d. as $\mathcal{CN}(0,1)$. In order to estimate the channel, a pilot sequence $\bar{\mathbf{X}}_{\mathrm{t}}\in \mathbb{C}^{K\times T_\mathrm{t}}$ of length $T_\mathrm{t}$ is used to generate the training data
\begin{equation}
\bar{\mathbf{Y}}_{\mathrm{t}} = \operatorname{sign}\left(\bar{\mathbf{H}}\bar{\mathbf{X}}_{\mathrm{t}}+\bar{\mathbf{Z}}_{\mathrm{t}}\right).
\label{eq_rx_training_seq}
\end{equation}
For convenience in later derivations, we convert the notation in~\eqref{eq_rx_training_seq} to the real domain as
\begin{equation}
\mathbf{Y}_{\mathrm{t}} = \operatorname{sign}\left(\mathbf{H}_{\mathrm{t}}\mathbf{X}_{\mathrm{t}} + \mathbf{Z}_{\mathrm{t}}\right),
\end{equation}
where 
\begin{align}
\mathbf{Y}_{\mathrm{t}} &= \left[\Re\{\bar{\mathbf{Y}}_\mathrm{t}\}, \Im \{\bar{\mathbf{Y}}_\mathrm{t}\}\right] = [\mathbf{y}_{\mathrm{t},1},\mathbf{y}_{\mathrm{t},2},\ldots,\mathbf{y}_{\mathrm{t},N}]^T, \label{eq_training_received_matrix_inReal}\\
\mathbf{H}_{\mathrm{t}} &= \left[\Re\{\bar{\mathbf{H}}\}, \Im \{\bar{\mathbf{H}}\}\right] = [\mathbf{h}_{\mathrm{t},1},\mathbf{h}_{\mathrm{t},2},\ldots,\mathbf{h}_{\mathrm{t},N}]^T, \label{eq_training_channel_matrix_inReal}\\
\mathbf{Z}_{\mathrm{t}} &= \left[\Re\{\bar{\mathbf{Z}}_{\mathrm{t}}\}, \Im \{\bar{\mathbf{Z}}_{\mathrm{t}}\}\right] = [\mathbf{z}_{\mathrm{t},1},\mathbf{z}_{\mathrm{t},2},\ldots,\mathbf{z}_{\mathrm{t},N}]^T, \text{ and} \label{eq_training_noise_matrix_inReal}\\
\mathbf{X}_{\mathrm{t}} &= \begin{bmatrix}
\Re \{\bar{\mathbf{X}}_{\mathrm{t}}\} & \Im \{\bar{\mathbf{X}}_{\mathrm{t}}\}\\
-\Im \{\bar{\mathbf{X}}_{\mathrm{t}}\} & \Re \{\bar{\mathbf{X}}_{\mathrm{t}}\}
\end{bmatrix} = [\mathbf{x}_{\mathrm{t},1},\mathbf{x}_{\mathrm{t},2},\ldots,\mathbf{x}_{\mathrm{t},2T_\mathrm{t}}]. \label{eq_training_matrix_inReal}
\end{align}
Note that $\mathbf{y}_{\mathrm{t},i}^T \in \{\pm1\}^{1\times 2T_\mathrm{t}}$, $\mathbf{h}_{\mathrm{t},i}^T\in \mathbb{R}^{1\times 2K}$, and $\mathbf{z}_{\mathrm{t},i}^T\in \mathbb{R}^{1\times 2T_\mathrm{t}}$ with $i\in \{1,2,\ldots,N\}$ represent the $i^\mathrm{th}$ rows of $\mathbf{Y}_{\mathrm{t}}$, $\mathbf{H}_{\mathrm{t}}$, and $\mathbf{Z}_{\mathrm{t}}$, respectively. However, $\mathbf{x}_{\mathrm{t},n} \in \mathbb{R}^{2K\times 1}$ with  $n \in \{1,2,\ldots,2T_\mathrm{t}\}$ is the $n^\mathrm{th}$ column of $\mathbf{X}_{\mathrm{t}}$.

It can be seen from~\eqref{eq_training_channel_matrix_inReal} that estimating $\{\mathbf{h}_{\mathrm{t},i}\}_{i=1,2,\ldots,N}$ is equivalent to estimating $\bar{\mathbf{H}}$. Here, we formulate the channel estimation problem in terms of $\mathbf{h}_{\mathrm{t},i}$. Let
\begin{align*}
\mathbf{y}_{\mathrm{t},i} &= [y_{\mathrm{t},i,1},y_{\mathrm{t},i,2},\ldots,y_{\mathrm{t},i,2T_{\mathrm{t}}}]^T \text{ and}\\
\mathbf{z}_{\mathrm{t},i} &= [z_{\mathrm{t},i,1},z_{\mathrm{t},i,2},\ldots,z_{\mathrm{t},i,2T_{\mathrm{t}}}]^T,
\end{align*}
then we have 
\begin{equation}
y_{\mathrm{t},i,n} = \operatorname{sign}\left(\mathbf{h}_{\mathrm{t},i}^T\mathbf{x}_{\mathrm{t},n}+z_{\mathrm{t},i,n}\right).
\label{eq_chanEst_as_binary_classification}
\end{equation}

We stress that the estimation of $\mathbf{h}_{\mathrm{t},i}$ in~\eqref{eq_chanEst_as_binary_classification} can be interpreted as an SVM binary classification problem. More specifically, $\left\{\mathbf{x}_{\mathrm{t},n},y_{\mathrm{t},i,n}\right\}_{n=1,\ldots,2T_\mathrm{t}}$ plays the role of the training data set $\mathcal{D}$. The channel $\mathbf{h}_{\mathrm{t},i}$ acts as the weight vector and $z_{\mathrm{t},i,n}$ can be viewed as the bias. Hence, we can follow the SVM classification formulation to estimate $\mathbf{h}_{\mathrm{t},i}$ by solving the following optimization problem:
\begin{equation}
\begin{aligned}
&\underset{\{\mathbf{h}_{\mathrm{t},i},\xi_n\}}{\operatorname{minimize}}
& & \frac{1}{2}\|\mathbf{h}_{\mathrm{t},i}\|^2+C\sum_{n=1}^{2T_\mathrm{t}}\xi_n\\
& \operatorname{subject\ to}
& & y_{\mathrm{t},i,n}\mathbf{h}_{\mathrm{t},i}^T\mathbf{x}_{\mathrm{t},n}\geq 1 - \xi_n,\\
& & & \xi_n \geq 0, \quad n = 1, 2, \ldots, 2T_\mathrm{t}.
\end{aligned}
\label{eq_SVM_chEst_formulation}
\end{equation}
Here, the bias is discarded because the $\{z_{\mathrm{t},i,n}\}$ are random noise with zero mean. In addition, at infinite SNR,~\eqref{eq_chanEst_as_binary_classification} becomes $y_{\mathrm{t},i,n} = \operatorname{sign}\big(\mathbf{h}_{\mathrm{t},i}^T\mathbf{x}_{\mathrm{t},n}\big)$, which has no bias. It should be noted that~\eqref{eq_SVM_chEst_formulation} only depends on a single index $i$, and so its solution is the estimate for the $i^\mathrm{th}$ row of the channel matrix $\bar{\mathbf{H}}$, i.e., the channel vector from the $K$ users to the $i$th receive antenna. This means we have $N$ separate optimization problems of the same form~\eqref{eq_SVM_chEst_formulation}, which is an advantage of the proposed SVM-based method since these $N$ optimization problems can be solved in parallel. 

Let $\tilde{\mathbf{h}}_{\mathrm{t},i}$ denote the solution of~\eqref{eq_SVM_chEst_formulation}. This solution provides an estimate of the channel ``direction'', but the magnitude of $\tilde{\mathbf{h}}_{\mathrm{t},i}$ is determined by the definition of the SVM margin, which in turn defines the inequality constraints in~\eqref{eq_SVM_chEst_formulation}. In fact, the instantaneous magnitude of $\mathbf{h}_{\mathrm{t},i}$ is not identifiable~\cite{Rao2019Channel} since $\alpha \mathbf{h}_{\mathrm{t},i}$ for any $\alpha > 0$ will produce the same data set $\{y_{t,i,n}\}$:
\begin{equation*}
y_{\mathrm{t},i,n} = \operatorname{sign}\big(\mathbf{h}_{\mathrm{t},i}^T\mathbf{x}_{\mathrm{t},n}\big) = \operatorname{sign}\big(\alpha\mathbf{h}_{\mathrm{t},i}^T\mathbf{x}_{\mathrm{t},n}\big),\ \text{with}\ \alpha >0.
\end{equation*}
Since in our model we assume that the $2K$ elements of $\mathbf{h}_{\mathrm{t},i}$ are each independent with variance $1/2$, we will scale the SVM solution so that the corresponding channel estimate has a squared norm of $K$:
\begin{equation}
\hat{\mathbf{h}}_{\mathrm{t},i} = \frac{\sqrt{K}\tilde{\mathbf{h}}_{\mathrm{t},i}}{\|\tilde{\mathbf{h}}_{\mathrm{t},i}\|}.
\label{eq_normalization_ChanEst}
\end{equation}
We have found that this choice for the scaling provides the best estimation accuracy.

\emph{Remark 1:} The soft-SVM method in~\cite{Kang2017Power} does not maximize the margin, but instead calculates $\mathbf{h}_{\mathrm{t},i}$ such that the condition $y_{\mathrm{t},i,n}\mathbf{h}_{\mathrm{t},i}^T\mathbf{x}_{\mathrm{t},n} > 0$ is satisfied for as many $n$ as possible. However, due to the noise component $z_{\text{t},i,n}$, the condition  $y_{\mathrm{t},i,n}\mathbf{h}_{\mathrm{t},i}^T\mathbf{x}_{\mathrm{t},n} > 0$ may not be satisfied even with the true channel vector $\mathbf{h}_{\mathrm{t},i}$. Our proposed method exploits the original idea of SVM by maximizing the margin achieved by the linear discriminator. The introduction of the slack variables in the problem circumvents the strict constraint $y_{\mathrm{t},i,n}\mathbf{h}_{\mathrm{t},i}^T\mathbf{x}_{\mathrm{t},n} > 0$.

\emph{Remark 2:} Without slack variables, the problem in~\eqref{eq_SVM_chEst_formulation}
\begin{equation}
\begin{aligned}
&\underset{\{\mathbf{h}_{\mathrm{t},i}\}}{\operatorname{minimize}}
& & \frac{1}{2}\|\mathbf{h}_{\mathrm{t},i}\|^2\\
& \operatorname{subject\ to}
& & y_{\mathrm{t},i,n}\mathbf{h}_{\mathrm{t},i}^T\mathbf{x}_{\mathrm{t},n}\geq 1, \quad n = 1, 2, \ldots, 2T_\mathrm{t},
\end{aligned}
\label{eq_SVM_chEst_original_formulation}
\end{equation}
is similar to the form in~\eqref{eq_SVM_problem}. For $\mathbf{h}_{\mathrm{t},i} \sim \mathcal{N}(\mathbf{0},\mathbf{I})$ we have
\begin{equation}
p(\mathbf{h}_{\mathrm{t},i}) = \frac{1}{\sqrt{(2\pi)^{2K}}}\exp \left\{-\frac{1}{2}\|\mathbf{h}_{\mathrm{t},i}\|^2\right\},
\end{equation}
and hence the optimization problem in~\eqref{eq_SVM_chEst_original_formulation} can be read as maximizing the pdf of $\mathbf{h}_{\mathrm{t},i}$ subject to the constraints $y_{\mathrm{t},i,n}\mathbf{h}_{\mathrm{t},i}^T\mathbf{x}_{\mathrm{t},n}\geq 1$ for $n = 1,2,\ldots,2T_\mathrm{t}$. Thus, the SVM approach can be interpreted as finding the channel $\mathbf{h}_{\mathrm{t},i}$ that attains the highest likelihood under the constraints realized by the measured data. We will use this observation next to modify the SVM-based channel estimator when the channel is spatially correlated. Note that the work in~\cite{Kang2017Power} only considers uncorrelated channels.
 
\subsubsection{Spatially Correlated Channels}
We let $\bar{\mathbf{H}} = [\bar{\mathbf{h}}_1, \ldots, \bar{\mathbf{h}}_K]$, and so $\bar{\mathbf{h}}_k \in \mathbb{C}^{N\times1}$ is the $k^\mathrm{th}$ column of $\bar{\mathbf{H}}$. Here, we assume that the elements of $\bar{\mathbf{h}}_k$ are correlated, or in other words that the channels associated with different antennas are correlated. Let $\bar{\mathbf{h}}_k \sim \mathcal{CN}(\mathbf{0},\bar{\mathbf{C}}_k)$ and $\bar{\mathbf{h}} = \operatorname{vec}(\bar{\mathbf{H}})$, then $\bar{\mathbf{h}} \sim \mathcal{CN}(\mathbf{0},\bar{\mathbf{C}})$ where $\bar{\mathbf{C}} = \operatorname{blockdiag}(\bar{\mathbf{C}}_1,\bar{\mathbf{C}}_2,\ldots,\bar{\mathbf{C}}_K)$. The pdf of $\bar{\mathbf{h}}$ is
\begin{align}
p(\bar{\mathbf{h}}) &= \frac{1}{\pi^{KN}\sqrt{\det(\bar{\mathbf{C}})}}\exp \left\{-\bar{\mathbf{h}}^H\bar{\mathbf{C}}^{-1}\bar{\mathbf{h}}\right\}\label{eq_joint_pdf_correlated_channel_ver1}\\
&= \frac{1}{\pi^{KN}\sqrt{\det(\bar{\mathbf{C}})}}\exp \left\{-\sum_{k=1}^K\bar{\mathbf{h}}_k^H\bar{\mathbf{C}}_k^{-1}\bar{\mathbf{h}}_k\right\}.
\label{eq_joint_pdf_correlated_channel}
\end{align}
The exponent term in~\eqref{eq_joint_pdf_correlated_channel_ver1} becomes a sum in~\eqref{eq_joint_pdf_correlated_channel} because $\bar{\mathbf{C}}$ is a block diagonal matrix, whose main-diagonal blocks are $\bar{\mathbf{C}}_1,\bar{\mathbf{C}}_2,\ldots,\bar{\mathbf{C}}_K$. Letting
\begin{equation*}
\mathbf{h}_k = \begin{bmatrix}
\Re\{\bar{\mathbf{h}}_k\}\\\Im\{\bar{\mathbf{h}}_k\}
\end{bmatrix} \text{ and } \mathbf{C}_k = \begin{bmatrix}
\Re\{\bar{\mathbf{C}}_k\} & -\Im\{\bar{\mathbf{C}}_k\}\\
\Im\{\bar{\mathbf{C}}_k\} & \Re\{\bar{\mathbf{C}}_k\}
\end{bmatrix},
\end{equation*}
the exponent term in~\eqref{eq_joint_pdf_correlated_channel} can be rewritten as $\sum_{k=1}^K\mathbf{h}_k^T\mathbf{C}_k^{-1} \mathbf{h}_k$.

To maximize the likelihood of $\bar{\mathbf{h}}$ subject to the constraints $y_{\mathrm{t},i,n}\mathbf{h}_{\mathrm{t},i}^T\mathbf{x}_{\mathrm{t},n}\geq 1$ with $i=1,2,\ldots,N$ and $n = 1,2,\ldots,2T_\mathrm{t}$, we follow the intuition in~\eqref{eq_SVM_chEst_original_formulation} to formulate the following optimization problem:
\begin{equation}
\begin{aligned}
&\underset{\{\bar{\mathbf{H}}\}}{\operatorname{minimize}}
& & \frac{1}{2} \sum_{k=1}^{K} \|\mathbf{h}_k^T\mathbf{C}^{-1}_k\mathbf{h}_k\|^2\\
& \operatorname{subject\ to}
& & y_{\mathrm{t},i,n}\mathbf{h}_{\mathrm{t},i}^T\mathbf{x}_{\mathrm{t},n}\geq 1,\\
& & & i = 1, 2, \ldots, N \text{ and }n = 1, 2, \ldots, 2T_\mathrm{t}.
\end{aligned}
\label{eq_SVM_correlated_chEst_original_formulation}
\end{equation}

In the above optimization problem, it is important to note that $\mathbf{h}_k \in \mathbb{R}^{2N\times 1}$ represents the $k^\mathrm{th}$ column of $\bar{\mathbf{H}}$, but $\mathbf{h}_{\mathrm{t},i}^T$ represents the $i^{\mathrm{th}}$ row of $\bar{\mathbf{H}}$. This means the objective function of~\eqref{eq_SVM_correlated_chEst_original_formulation} depends on the columns of $\bar{\mathbf{H}}$, but the constraints depend on the rows of $\bar{\mathbf{H}}$. Therefore, we cannot decompose~\eqref{eq_SVM_correlated_chEst_original_formulation} into smaller independent problems. In other words, the whole channel matrix $\bar{\mathbf{H}}$ has to be jointly estimated.

We note that the margin $\mathbf{h}_k^T\mathbf{C}_k^{-1}\mathbf{h}_k$ in \eqref{eq_SVM_correlated_chEst_original_formulation} is measured using the Mahalanobis distance~\cite{mahalanobis1936generalized}  rather than the Euclidean metric used in the standard SVM approach. The optimization problem in~\eqref{eq_SVM_correlated_chEst_original_formulation} can also be generalized by including slack variables as
\begin{equation}
\begin{aligned}
&\underset{\{\bar{\mathbf{H}},\xi_{i,n}\}}{\operatorname{minimize}}
& & \frac{1}{2} \sum_{k=1}^{K} \|\mathbf{h}_k^T\mathbf{C}^{-1}_k\mathbf{h}_k\|^2+C\sum_{i=1}^N\sum_{n=1}^{2T_\mathrm{t}}\xi_{i,n}\\
& \operatorname{subject\ to}
& & y_{\mathrm{t},i,n}\mathbf{h}_{\mathrm{t},i}^T\mathbf{x}_{\mathrm{t},n}\geq 1 - \xi_{i,n} \text{ with }\xi_{i,n} \geq 0,\\
& & & i = 1, 2, \ldots, N \text{ and }n = 1, 2, \ldots, 2T_\mathrm{t}.
\end{aligned}
\label{eq_SVM_correlated_chEst_formulation}
\end{equation}

Although the form of the objective function in~\eqref{eq_SVM_correlated_chEst_formulation} is different from that in conventional SVM problems,~\eqref{eq_SVM_correlated_chEst_formulation} can still be solved efficiently since it is a convex optimization problem. Let $\tilde{\mathbf{H}}$ be the solution of~\eqref{eq_SVM_correlated_chEst_formulation}, then the channel estimate $\hat{\mathbf{H}}$ is defined as
\begin{equation*}
	\hat{\mathbf{H}} = \frac{\sqrt{KN}\tilde{\mathbf{H}}}{\|\tilde{\mathbf{H}}\|_\mathrm{F}},
\end{equation*}
where $\|\cdot\|_\mathrm{F}$ denotes the Frobenius norm. This normalization step is similar to that for the case of uncorrelated channels, except a different coefficient $\sqrt{KN}$ is used since we jointly estimate the whole channel matrix and $\mathbb{E}[\|\bar{\mathbf{H}}\|_\mathrm{F}] = \sqrt{KN}$.
\subsection{Proposed Two-Stage SVM-based Data Detection}
In this section, we propose a two-stage SVM-based method for data detection with $1$-bit ADCs. We first formulate the data detection as an SVM problem. A second stage is then employed to refine the solution from the first stage. Let $\bar{\mathbf{X}}_\mathrm{d} = [\bar{\mathbf{x}}_{\mathrm{d},1}, \bar{\mathbf{x}}_{\mathrm{d},2}, \ldots, \bar{\mathbf{x}}_{\mathrm{d},T_\mathrm{d}}] \in \mathbb{C}^{K\times T_\mathrm{d}}$ be the transmitted data sequence of length $T_\mathrm{d}$. The received data signal is given as
\begin{equation}
\bar{\mathbf{Y}}_\mathrm{d} = \operatorname{sign}\left(\bar{\mathbf{H}}\bar{\mathbf{X}}_\mathrm{d} + \bar{\mathbf{Z}}_\mathrm{d}\right).
\label{eq_data_received_signal}
\end{equation}
The above equation is also converted to the real domain as
\begin{equation}
\mathbf{Y}_\mathrm{d} = \operatorname{sign}\left(\mathbf{H}_\mathrm{d}\mathbf{X}_\mathrm{d} + \mathbf{Z}_\mathrm{d}\right)
\end{equation}
where
\begin{align}
\mathbf{Y}_{\mathrm{d}} &= \begin{bmatrix}
\Re\{\bar{\mathbf{Y}}_\mathrm{d}\} \\ \Im \{\bar{\mathbf{Y}}_\mathrm{d}\}
\end{bmatrix}= [\mathbf{y}_{\mathrm{d},1},\mathbf{y}_{\mathrm{d},2},\ldots,\mathbf{y}_{\mathrm{d},T_\mathrm{d}}],\label{eq_data_rxSig_matrix_inReal}\\
\mathbf{X}_{\mathrm{d}} &= \begin{bmatrix}
\Re\{\bar{\mathbf{X}}_\mathrm{d}\}\\ \Im \{\bar{\mathbf{X}}_\mathrm{d}\}
\end{bmatrix} = [\mathbf{x}_{\mathrm{d},1},\mathbf{x}_{\mathrm{d},2},\ldots,\mathbf{x}_{\mathrm{d},T_\mathrm{d}}],\label{eq_data_matrix_inReal}\\
\mathbf{Z}_{\mathrm{d}} &= \begin{bmatrix}
\Re\{\bar{\mathbf{Z}}_{\mathrm{d}}\} \\ \Im \{\bar{\mathbf{Z}}_{\mathrm{d}}\}
\end{bmatrix} = [\mathbf{z}_{\mathrm{d},1},\mathbf{z}_{\mathrm{d},2},\ldots,\mathbf{z}_{\mathrm{d},T_\mathrm{d}}], \label{eq_data_noise_matrix_inReal}\text{ and}\\
\mathbf{H}_{\mathrm{d}} &= \begin{bmatrix}
\Re \{\bar{\mathbf{H}}\} & -\Im \{\bar{\mathbf{H}}\}\\
\Im \{\bar{\mathbf{H}}\} &  \Re \{\bar{\mathbf{H}}\}
\end{bmatrix} = [\mathbf{h}_{\mathrm{d},1},\mathbf{h}_{\mathrm{d},2},\ldots,\mathbf{h}_{\mathrm{d},2N}]^T.\label{eq_data_channel_matrix_inReal}
\end{align}
Here, $\mathbf{y}_{\mathrm{d},m} \in \{\pm1\}^{2N\times1}$, $\mathbf{x}_{\mathrm{d},m}\in \mathbb{R}^{2K\times1}$, and $\mathbf{z}_{\mathrm{d},m}\in \mathbb{R}^{2N\times1}$ with $m\in \{1,2,\ldots,T_\mathrm{d}\}$ are the $m^\mathrm{th}$ columns of $\mathbf{Y}_{\mathrm{d}}$, $\mathbf{X}_{\mathrm{d}}$, and $\mathbf{Z}_{\mathrm{d}}$, respectively. However, $\mathbf{h}_{\mathrm{d},i'}^T \in \mathbb{R}^{1\times2K}$ with $i' \in \{1,2,\ldots,2N\}$ represents the ${i'}^{\mathrm{th}}$ row of $\mathbf{H}_{\mathrm{d}}$.

It can be noted that the real and imaginary parts in~\eqref{eq_training_received_matrix_inReal}--\eqref{eq_training_matrix_inReal} are stacked side-by-side, but they are stacked on top of each other in~\eqref{eq_data_rxSig_matrix_inReal}--\eqref{eq_data_channel_matrix_inReal}. This is due to the exchange in the role of the channel and the data matrices. In the formulation for channel estimation in~\eqref{eq_training_received_matrix_inReal}--\eqref{eq_training_matrix_inReal}, each row of the channel matrix is treated as the weight vector and the columns of the pilot data matrix are used as the training data points. On the other hand, the data detection formulation in~\eqref{eq_data_rxSig_matrix_inReal}--\eqref{eq_data_channel_matrix_inReal} treats each column of the to-be-decoded data matrix as the weight vector and the rows of the channel matrix as the training data points.

It should also be noted that the pilot sequence and the data sequence are assumed to experience the same block-fading channel. Although the two channel matrices $\mathbf{H}_{\mathrm{t}}$ in~\eqref{eq_training_channel_matrix_inReal} and $\mathbf{H}_{\mathrm{d}}$ in~\eqref{eq_data_channel_matrix_inReal} are constructed differently, they still depend on the same channel $\bar{\mathbf{H}}$. Let
\begin{align*}
\mathbf{y}_{\mathrm{d},m} &= [y_{\mathrm{d},m,1},y_{\mathrm{d},m,2},\ldots,y_{\mathrm{d},m,2N}]^T \text{ and}\\
\mathbf{z}_{\mathrm{d},m} &= [z_{\mathrm{d},m,1},z_{\mathrm{d},m,2},\ldots,z_{\mathrm{d},m,2N}]^T,
\end{align*}
then we have 
\begin{equation}
y_{\mathrm{d},m,i'} = \operatorname{sign}\Big(\mathbf{h}_{\mathrm{d},i'}^T\mathbf{x}_{\mathrm{d},m}+z_{\mathrm{d},m,i'}\Big).
\label{eq_dataDetection_as_binary_classification}
\end{equation}

It is observed that the estimation of $\mathbf{x}_{\mathrm{d},m}$ can also be interpreted as an SVM binary classification problem. More specifically, we can treat $\mathbf{x}_{\mathrm{d},m}$ as the weight vector and the set $\{\hat{\mathbf{h}}_{\mathrm{d},i'}, y_{\mathrm{d},m,i'}\}_{i'=1,\ldots,2N}$ as the training set, where $\hat{\mathbf{h}}_{\mathrm{d},i'}$ is the channel estimate of $\mathbf{h}_{\mathrm{d},i'}$ obtained as explained above. The following optimization problem provides the first-stage in finding $\mathbf{x}_{\mathrm{d},m}$:
\begin{equation}
\begin{aligned}
&\underset{\{\mathbf{x}_{\mathrm{d},m},\xi_{i'}\}}{\operatorname{minimize}}
& & \frac{1}{2}\|\mathbf{x}_{\mathrm{d},m}\|^2+C\sum_{i=1}^{2N}\xi_{i'}\\
& \operatorname{subject\ to}
& & y_{\mathrm{d},m,i'}\mathbf{x}_{\mathrm{d},m}^T\hat{\mathbf{h}}_{\mathrm{d},i'}\geq 1 - \xi_{i'},\\
& & & \xi_{i'} \geq 0, \quad i' = 1, 2, \ldots, 2N,
\end{aligned}
\label{eq_SVM_data_detection_formulation}
\end{equation}
where the bias is discarded as in the channel estimation problem. Let $\tilde{\mathbf{x}}_{\mathrm{d},m}$ denote the solution of~\eqref{eq_SVM_data_detection_formulation} and let $\grave{\mathbf{x}}_{\mathrm{d},m}$ be the normalized version of $\tilde{\mathbf{x}}_{\mathrm{d},m}$ as
\begin{equation}
\grave{\mathbf{x}}_{\mathrm{d},m} = \frac{\sqrt{K}\tilde{\mathbf{x}}_{\mathrm{d},m}}{\|\tilde{\mathbf{x}}_{\mathrm{d},m}\|}.
\end{equation}                                                        
This normalization step is also used in~\cite{choi2016near} in order to make the power of the estimated signal equal the power of the transmitted signal.

Let $\grave{\mathbf{x}}_{\mathrm{d},m} = [\grave{x}_{\mathrm{d},m,1}, \ldots,\grave{x}_{\mathrm{d},m,2K}]^T$, and define the first-stage detected data vector $\check{\mathbf{x}}_{\mathrm{d},m} = [\check{x}_{\mathrm{d},m,1}, \ldots,\check{x}_{\mathrm{d},m,K}]^T$ obtained using symbol-by-symbol detection as
\begin{equation}
\check{x}_{\mathrm{d},m,k} = \argmin_{x\in \mathcal{M}} \left|(\grave{x}_{\mathrm{d},m,k}+j\grave{x}_{\mathrm{d},m,k+K})-x\right|,
\label{eq_SVM_stage1_sym-by-sym_detection}
\end{equation}
where $k \in \mathcal{K}$ and $\mathcal{M}$ represents the signal constellation (e.g., QPSK or $16$-QAM). The solution to~\eqref{eq_SVM_stage1_sym-by-sym_detection} is referred to as the stage 1 solution. To further improve the detection performance, a simple but efficient second detection stage is proposed as follows.

First, a candidate set $\mathcal{X}_k$ for each $\bar{x}_{\mathrm{d},m,k}$ is created using $\check{x}_{\mathrm{d},m,k}$ and $\grave{x}_{\mathrm{d},m,k}+j\grave{x}_{\mathrm{d},m,k+K}$ as
\begin{equation}
\mathcal{X}_{k} = \left\{\acute{x}\in\mathcal{M} \bigg \lvert \frac{|(\grave{x}_{\mathrm{d},m,k}+j\grave{x}_{\mathrm{d},m,k+K}) - \acute{x}|}{|(\grave{x}_{\mathrm{d},m,k}+j\grave{x}_{\mathrm{d},m,k+K})-\check{x}_{\mathrm{d},m,k}|} < \gamma \right\}
\end{equation}
where $\gamma \geq 1$ is a parameter that controls the size of $\mathcal{X}_k$. Then the candidate set $\mathcal{X}$ for $\mathbf{x}_{\mathrm{d},m}$ is obtained as
\begin{equation}
\mathcal{X} = \left\{[\acute{x}_1, \acute{x}_2, \ldots, \acute{x}_K]^T \mid \acute{x}_k \in \mathcal{X}_k, \forall k \in \mathcal{K}\right\}.
\end{equation}

The above candidate set formation was introduced in~\cite{choi2016near}. However, the detected data vector in~\cite{choi2016near} is obtained by searching over $\mathcal{X}$ using the ML criterion, and the resulting performance is susceptible to imperfect CSI at high SNRs. This susceptibility has been reported via numerical results in~\cite{Jeon2018supervised} and~\cite{nguyen2019supervised}, but no justification was given. We provide an explanation for this issue in Appendix~\ref{sec_appendA}. To deal with the issue, we adopt here a different criterion referred to as \textit{minimum weighted Hamming distance}~\cite{Jeon2018One}. Suppose that $\mathcal{X} = \{\acute{\mathbf{x}}_1, \acute{\mathbf{x}}_2, \ldots, \acute{\mathbf{x}}_{|\mathcal{X}|}\}$ and let $\dot{\mathbf{x}}_l = [\Re\{\acute{\mathbf{x}}_l\}^T, \Im\{\acute{\mathbf{x}}_l\}^T]^T$ with $l \in \{1,2,\ldots,|\mathcal{X}|\}$. The second-stage detected data vector $\hat{\mathbf{x}}_{\mathrm{d},m}$ is defined as $\hat{\mathbf{x}}_{\mathrm{d},m} = \acute{\mathbf{x}}_{\hat{l}}$ where
\begin{equation}
\hat{l} = \argmin_{l \in \{1,\ldots,|\mathcal{X}|\}} d_\mathrm{w}\left(\mathbf{y}_{\mathrm{d},m},\operatorname{sign}(\hat{\mathbf{H}}_{\mathrm{d}}\dot{\mathbf{x}}_l)\right).
\label{eq_minimum_weighted_Hamming_distance}
\end{equation}
Here, $\hat{\mathbf{H}}_{\mathrm{d}}$ is the channel estimate of $\mathbf{H}_{\mathrm{d}}$ and $d_\mathrm{w}(\cdot,\cdot)$ is the weighted Hamming distance defined in~\cite{Jeon2018One}.

The minimum weighted Hamming distance criterion above was shown to be statistically efficient~\cite{Jeon2018One}. However, the OSD method proposed in~\cite{Jeon2018One} requires a preprocessing stage whose computational complexity is proportional to $2^{N_\mathrm{s}}|\mathcal{M}|^{N_\mathrm{t}}$ for each channel realization. Here $N_\mathrm{s} = 2N/G$ where $G\geq1$ is an integer. The exponential computational complexity of OSD is a significant drawback in large-scale system implementation. The proposed SVM-based data detection method in this paper can address this complexity issue since the optimization problem~\eqref{eq_SVM_data_detection_formulation} can be solved by very efficient algorithms~\cite{joachims2006training,keerthi2005modified,bottou2007support}.

\subsection{Proposed SVM-based Joint CE-DD}
In $1$-bit ADC systems, the channel estimation accuracy can be improved by increasing the length of the pilot training sequence, but not necessarily by increasing the SNR~\cite{li2017channel}. For this reason, we propose here an SVM-based joint CE-DD method to effectively improve the channel estimate without lengthening the pilot training sequence. The idea is to use the detected data vectors from the two-stage SVM-based method together with the pilot data vectors to obtain a refined channel estimate and then use this refined channel estimate to improve the data detection performance. 

Let $\hat{\mathbf{X}}_\mathrm{d}$ be the detected version of $\bar{\mathbf{X}}_\mathrm{d}$ using the proposed two-stage data detection method and let
\begin{align}
\hat{\mathbf{X}}_{\mathrm{d2}} &= \begin{bmatrix}
\Re \{\hat{\mathbf{X}}_{\mathrm{d}}\} & \Im \{\hat{\mathbf{X}}_{\mathrm{d}}\}\\
-\Im \{\hat{\mathbf{X}}_{\mathrm{d}}\} & \Re \{\hat{\mathbf{X}}_{\mathrm{d}}\}
\end{bmatrix} = [\hat{\mathbf{x}}_{\mathrm{d2},1},\ldots,\hat{\mathbf{x}}_{\mathrm{d2},2T_\mathrm{d}}],\\
\mathbf{Y}_{\mathrm{d2}} &= \left[\Re\{\bar{\mathbf{Y}}_\mathrm{d}\}, \Im \{\bar{\mathbf{Y}}_\mathrm{d}\}\right] = [\mathbf{y}_{\mathrm{d2},1},\ldots,\mathbf{y}_{\mathrm{d2},N}]^T,
\end{align}
where $\mathbf{y}_{\mathrm{d2},i} = [y_{\mathrm{d2},i,1}, y_{\mathrm{d2},i,2}, \ldots, y_{\mathrm{d2},i,2T_\mathrm{d}}]^T$, $i = 1,\ldots,N$. The channel estimate can be refined by solving the following optimization problem:
\begin{equation}
\begin{aligned}
&\underset{\{\mathbf{h}_{\mathrm{t},i},\xi_{\mathrm{t},n},\xi_{\mathrm{d},m}\}}{\operatorname{minimize}}
& & \frac{1}{2}\|\mathbf{h}_{\mathrm{t},i}\|^2+C\left(\sum_{n=1}^{2T_\mathrm{t}}\xi_{\mathrm{t},n}+\sum_{m=1}^{2T_\mathrm{d}}\xi_{\mathrm{d},m}\right)\\
& \operatorname{subject\ to}
& & y_{\mathrm{t},i,n}\mathbf{h}_{\mathrm{t},i}^T\mathbf{x}_{\mathrm{t},n}\geq 1 - \xi_{\mathrm{t},n},\\
& & & y_{\mathrm{d2},i,m}\mathbf{h}_{\mathrm{t},i}^T\hat{\mathbf{x}}_{\mathrm{d2},m}\geq 1 - \xi_{\mathrm{d},m},\\
& & & \xi_{\mathrm{t},n} \geq 0, \quad n = 1, 2, \ldots, 2T_\mathrm{t},\\
& & & \xi_{\mathrm{d},m} \geq 0, \quad m = 1, 2, \ldots, 2T_\mathrm{d}.
\end{aligned}
\label{eq_SVM_chEst_formulation2}
\end{equation}

In the optimization problem above, we use two sets of slack variables $\{\xi_{\mathrm{t},n}\}$ and $\{\xi_{\mathrm{d},m}\}$, which correspond to the pilot sequence and the data sequence, respectively. This is just for notational convenience, as the two sets of slack variables play the same role. The refined channel estimate obtained by solving~\eqref{eq_SVM_chEst_formulation2} can now be used for data detection again in~\eqref{eq_SVM_data_detection_formulation} and~\eqref{eq_minimum_weighted_Hamming_distance}. Note that the channel estimate obtained by~\eqref{eq_SVM_chEst_formulation} can be used as the initial solution to~\eqref{eq_SVM_chEst_formulation2} so that the algorithm will more quickly converge to the optimal solution. Similarly, $\hat{\mathbf{X}}_{\mathrm{d}}$ can also be used as the initial solution when solving~\eqref{eq_SVM_data_detection_formulation} with the refined channel estimate. Numerical results in Section~\ref{sec_Numerical_Results} show that this strategy will hit a certain performance bound as $T_\mathrm{d}$ increases.

\section{Extension to OFDM systems with Frequency-Selective Fading Channels}
\label{sec_extension_to_wideband}
In this section, we develop SVM-based channel estimation and SVM-based data detection for OFDM systems with frequency-selective fading channels. Consider an uplink multiuser OFDM system with $N_\mathrm{c}$ subcarriers. Denote $\bar{\mathbf{x}}_k^{\mathrm{FD}} \in \mathbb{C}^{N_{\mathrm{c}}\times 1}$ as the  OFDM symbol from the $k^\mathrm{th}$ user in the frequency domain. Throughout the paper, we use the superscripts ``TD'' and ``FD'' to refer to Time Domain and Frequency Domain, respectively. A cyclic prefix (CP) of length $N_{\mathrm{cp}}$ is added and the number of channel taps $L$ is assumed to satisfy $L-1\leq N_{\mathrm{cp}} \leq N_{\mathrm{c}}$. It is assumed that $L$ is known. After removing the CP, the quantized received signal at the $i^{\text{th}}$ antenna in the time domain is given by
\begin{equation}
	\bar{\mathbf{y}}_{i}^{\mathrm{TD}} = \operatorname{sign}\left(\sum_{k=1}^{K}\bar{\mathbf{G}}_{i,k}^{\mathrm{TD}} \mathbf{F}^{H}\bar{\mathbf{x}}_k^{\mathrm{FD}} + \bar{\mathbf{z}}_i^{\mathrm{TD}}\right)
	\label{eq_OFDM_model}
\end{equation}
where $\mathbf{F}$ is the DFT matrix of size $N_\mathrm{c} \times N_\mathrm{c}$;  $\bar{\mathbf{G}}_{i,k}^{\mathrm{TD}}$ is a circulant matrix whose first column is  $\bar{\mathbf{g}}_{i,k}^{\mathrm{TD}} = [(\bar{\mathbf{h}}_{i,k}^{\mathrm{TD}})^T,0,\ldots,0]^T$; and $\bar{\mathbf{h}}_{i,k}^{\mathrm{TD}}$ is the channel vector of the $k^{\mathrm{th}}$ user containing the $L$ channel taps, which are assumed to be i.i.d. and distributed as $\mathcal{CN}(0,\frac{1}{L})$. We also assume block-fading channels where the first OFDM symbol is used for channel estimation and the other OFDM symbols in the block-fading interval are for data transmission. Thus, the problem of channel estimation and data detection are studied separately.
\subsection{Proposed SVM-based Channel Estimation in OFDM Systems with Frequency-Selective Fading Channels}
\label{subsec_chan_est_wideband}
Denote $\bar{\boldsymbol{\phi}}_k^{\mathrm{TD}} = \mathbf{F}^H\bar{\mathbf{x}}_k^{\mathrm{FD}}$ and the training matrix $\bar{\mathbf{\Phi}}_k^{\mathrm{TD}}$ as a circulant matrix with first column equal to $\bar{\boldsymbol{\phi}}_k^{\mathrm{TD}}$. We can reorganize the system model in \eqref{eq_OFDM_model} as follows:
\begin{align}
	\bar{\mathbf{y}}_{i}^{\mathrm{TD}} &= \operatorname{sign}\left(\sum_{k=1}^{K} \bar{\mathbf{\Phi}}_k^{\mathrm{TD}}\bar{\mathbf{g}}^{\mathrm{TD}}_{i,k} + \bar{\mathbf{z}}_i^{\mathrm{TD}}\right) \nonumber\\
	&=\operatorname{sign}\left(\sum_{k=1}^{K} \bar{\mathbf{\Phi}}_{k,L}^{\mathrm{TD}}\bar{\mathbf{h}}^{\mathrm{TD}}_{i,k} + \bar{\mathbf{z}}_i^{\mathrm{TD}}\right) \nonumber \\
	&=\operatorname{sign}\left(\bar{\boldsymbol{\Phi}}^{\mathrm{TD}}_L\bar{\mathbf{h}}^{\mathrm{TD}}_i + \bar{\mathbf{z}}_i^{\mathrm{TD}}\right) \label{eq_OFDM_chan_Est}
\end{align}
where $\bar{\boldsymbol{\Phi}}_{k,L}^{\mathrm{TD}}$ is the matrix corresponding to the first $L$ columns of $\bar{\mathbf{\Phi}}_{k}^{\mathrm{TD}}$, $\bar{\boldsymbol{\Phi}}^{\mathrm{TD}}_L = [\bar{\mathbf{\Phi}}_{1,L}^{\mathrm{TD}},\bar{\mathbf{\Phi}}_{2,L}^{\mathrm{TD}},\ldots,\bar{\mathbf{\Phi}}_{K,L}^{\mathrm{TD}}]$, and $\bar{\mathbf{h}}^{\mathrm{TD}}_i = [(\bar{\mathbf{h}}_{i,1}^{\mathrm{TD}})^T, (\bar{\mathbf{h}}^{\mathrm{TD}}_{i,2})^T,\ldots, (\bar{\mathbf{h}}^{\mathrm{TD}}_{i,K})^T]^T$.

We also convert \eqref{eq_OFDM_chan_Est} into the real domain as
\begin{equation}
	\mathbf{y}_{i}^{\mathrm{TD}} = \operatorname{sign}\left(\boldsymbol{\Phi}^{\mathrm{TD}}_L\mathbf{h}^{\mathrm{TD}}_i + \mathbf{z}_i^{\mathrm{TD}}\right)
\end{equation}
where
\begin{align*}
\mathbf{y}^\mathrm{TD}_i &= \left[\Re\{{\bar{\mathbf{y}}}^\mathrm{TD}_i\}^T, \Im \{\bar{\mathbf{y}}^\mathrm{TD}_i\}^T\right]^T,\\
\mathbf{h}^\mathrm{TD}_{i} &= \left[\Re\{\bar{\mathbf{h}}^\mathrm{TD}_i\}^T, \Im \{\bar{\mathbf{h}}^\mathrm{TD}_i\}^T\right]^T,\\
\mathbf{z}^\mathrm{TD}_{i} &= \left[\Re\{\bar{\mathbf{z}}^\mathrm{TD}_{i}\}^T, \Im \{\bar{\mathbf{z}}^\mathrm{TD}_{i}\}^T\right]^T, \text{ and}\\
\boldsymbol{\Phi}^{\mathrm{TD}}_L &= \begin{bmatrix}
\Re \{\bar{\boldsymbol{\Phi}}^{\mathrm{TD}}_L\} & -\Im \{\bar{\boldsymbol{\Phi}}^{\mathrm{TD}}_L\}\\
\Im \{\bar{\boldsymbol{\Phi}}^{\mathrm{TD}}_L\} & \Re \{\bar{\boldsymbol{\Phi}}^{\mathrm{TD}}_L\}
\end{bmatrix}.
\end{align*}
Denote $ \mathbf{y}^\mathrm{TD}_i = [y^{\mathrm{TD}}_{i,1},y^{\mathrm{TD}}_{i,2},\ldots,y^{\mathrm{TD}}_{i,2N_\mathrm{c}}]^T$ and $\boldsymbol{\Phi}^{\mathrm{TD}}_L=\big[(\boldsymbol{\phi}_1^{\mathrm{TD}})^T,(\boldsymbol{\phi}_2^{\mathrm{TD}})^T,\ldots,(\boldsymbol{\phi}_{2N_\mathrm{c}}^{\mathrm{TD}})^T\big]^T$, leading to the following SVM problem for estimating the OFDM channel using one-bit ADCs:
\begin{equation}
\begin{aligned}
&\underset{\{\mathbf{h}^{\mathrm{TD}}_{i},\xi_n\}}{\operatorname{minimize}}
& & \frac{1}{2}\|\mathbf{h}^{\mathrm{TD}}_{i}\|^2+C\sum_{n=1}^{2N_\mathrm{c}}\xi_{n}\\
& \operatorname{subject\ to}
& & y^{\mathrm{TD}}_{i,n}\left(\mathbf{h}^{\mathrm{TD}}_{i}\right)^T\boldsymbol{\phi}_{n}^{\mathrm{TD}}\geq 1 - \xi_{n},\\
& & & \xi_{n} \geq 0, \quad n = 1, 2, \ldots, 2N_\mathrm{c}.
\end{aligned}
\label{eq_SVM_OFDM_chEst_formulation}
\end{equation}
Denoting $\tilde{\mathbf{h}}^{\mathrm{TD}}_i$ as the solution of~\eqref{eq_SVM_OFDM_chEst_formulation}, then we estimate $\mathbf{h}^{\mathrm{TD}}_i$ as
\begin{equation}
	\hat{\mathbf{h}}^{\mathrm{TD}}_i = \frac{\sqrt{K}\tilde{\mathbf{h}}^{\mathrm{TD}}_i}{\|\tilde{\mathbf{h}}^{\mathrm{TD}}_i\|}.
\end{equation}

Frequency-selective channel estimation methods using one-bit ADCs have been previously proposed in~\cite{li2017channel,Mollen2017Uplink}, and \cite{balevi2019two} based on the Bussgang decomposition, Additive Quantization Noise Model (AQNM), and deep learning, respectively. The deep learning method in~\cite{balevi2019two} was shown to outperform the methods of~\cite{Mollen2017Uplink,balevi2019two} at low SNRs, but its performance tends to degrade as the SNR increases. In addition, the method in~\cite{balevi2019two} requires a training sequence that contains many OFDM symbols, which are required to be orthogonal between different users. In our proposed method, only one OFDM symbol is used in the training phase and all users send their training symbols concurrently.

\subsection{Proposed SVM-based Data Detection in OFDM Systems with Frequency-Selective Fading Channels}
\label{subsec_data_detect_wideband}
In this section, we describe how SVM can also be used for data detection in OFDM systems with frequency-selective fading channels. We can rewrite the received quantized vector in~\eqref{eq_OFDM_model} as
\begin{equation}
	\bar{\mathbf{y}}_{i}^{\mathrm{TD}} = \operatorname{sign}\left(\bar{\mathbf{G}}_{i}^{\mathrm{FD}}\bar{\mathbf{x}}^{\mathrm{FD}} + \bar{\mathbf{z}}_i^{\mathrm{TD}}\right)
\end{equation}
where $\bar{\mathbf{G}}_{i}^{\mathrm{FD}} = [\bar{\mathbf{G}}^{\mathrm{TD}}_{i,1} \mathbf{F}^{H},\ldots,\bar{\mathbf{G}}^{\mathrm{TD}}_{i,K} \mathbf{F}^{H}] \in \mathbb{C}^{N_\mathrm{c}\times N_\mathrm{c}K}$ and $\bar{\mathbf{x}}^{\mathrm{FD}} = [(\bar{\mathbf{x}}_1^{\mathrm{FD}})^T,\ldots,(\bar{\mathbf{x}}_K^{\mathrm{FD}})^T]^T$ is the transmitted symbol vector from the $K$ users over $N_\mathrm{c}$ subcarriers. By stacking all the received signal vectors $\left\{\bar{\mathbf{y}}_{i}^{\mathrm{TD}}\right\}_{i=1,\ldots,N}$ in a column vector, we have the following equation:
\begin{equation}
\bar{\mathbf{y}}^{\mathrm{TD}} = \operatorname{sign}\left(\bar{\mathbf{G}}^{\mathrm{FD}}\bar{\mathbf{x}}^{\mathrm{FD}} + \bar{\mathbf{z}}^{\mathrm{TD}}\right)
\label{eq_OFDM_data_detection}
\end{equation}
where $\bar{\mathbf{y}}^{\mathrm{TD}} = \big[(\bar{\mathbf{y}}_{1}^{\mathrm{TD}})^T,(\bar{\mathbf{y}}_{2}^{\mathrm{TD}})^T,\ldots,(\bar{\mathbf{y}}_{N}^{\mathrm{TD}})^T\big]^T$ and $\bar{\mathbf{G}}^{\mathrm{FD}} = \big[(\bar{\mathbf{G}}_{1}^{\mathrm{FD}})^T,(\bar{\mathbf{G}}_{2}^{\mathrm{FD}})^T,\ldots,(\bar{\mathbf{G}}_{N}^{\mathrm{FD}})^T\big]^T$. Let $\mathbf{y}^{\mathrm{TD}}$, $\mathbf{G}^{\mathrm{FD}}$, and $\mathbf{x}^{\mathrm{FD}}$ be the real-valued versions of $\bar{\mathbf{y}}^{\mathrm{TD}}$, $\bar{\mathbf{G}}^{\mathrm{FD}}$, and $\bar{\mathbf{x}}^{\mathrm{FD}}$, respectively. Converting \eqref{eq_OFDM_data_detection} to the real domain as in~\eqref{eq_data_rxSig_matrix_inReal}--\eqref{eq_data_channel_matrix_inReal}, we can formulate an SVM problem by treating the rows of $\mathbf{G}^{\mathrm{FD}}$ as the feature vectors, the elements of $\mathbf{y}^{\mathrm{TD}}$ as the binary indicators and $\mathbf{x}^{\mathrm{FD}}$ as the weight vector. The solution of the SVM problem then provides the detected data.

\section{Numerical Results}
\label{sec_Numerical_Results}
This section presents numerical results to show the superiority of the proposed methods against existing ones. For the simulations we set $C = 1$ and parameter $\gamma$ for the second stage of the SVM-based detection method as $\gamma = \min \left\{\frac{\rho}{10}+1.5,3\right\}$ for QPSK and $\gamma = \min \left\{\frac{\rho}{10}+1.3,1.5\right\}$ for $16$-QAM where $\rho$ is the SNR. The length of the block-fading interval is set to $500$ (i.e., $T_\mathrm{t} + T_\mathrm{d} = 500$) unless otherwise stated. For solving the proposed SVM-based channel estimation and data detection problems, we use the Scikit-learn machine learning library~\cite{scikit-learn}.
\begin{figure}[t!]
	\centering
	\includegraphics[width=\linewidth]{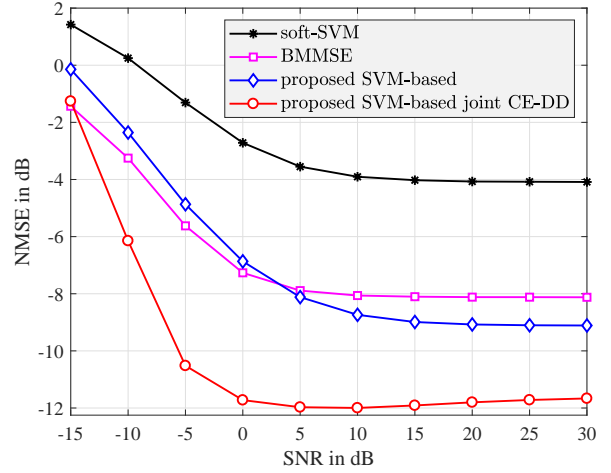}
	\caption{$\mathrm{NMSE}$ comparison between different channel estimators with $K = 4$, $N = 32$, and $T_\mathrm{t} = 20$.}
	\label{fig_ChanEst_Comparison_4K_32N}
\end{figure}
\begin{figure}[t!]
	\centering
	\includegraphics[width=\linewidth]{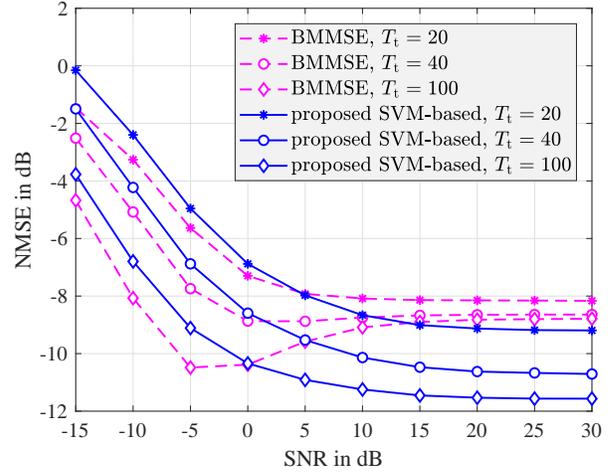}
	\caption{$\mathrm{NMSE}$ comparison between BMMSE and the proposed SVM-based channel estimator with $K = 4$, $N = 32$, and $T_\mathrm{t} \in \{20,40,100\}$.}
	\label{fig_ChanEst_4K_32N_different_taus.eps}
\end{figure}

Fig.~\ref{fig_ChanEst_Comparison_4K_32N} presents a performance comparison of different channel estimation methods in terms of $\mathrm{NMSE}$, defined here as $\mathrm{NMSE} = \mathbb{E}\big[\|\hat{\mathbf{H}}-\bar{\mathbf{H}}\|^2_\mathrm{F}\big]/(KN)$,
where $\hat{\mathbf{H}}$ is an estimate of the channel $\bar{\mathbf{H}}$. It can first be seen that the soft-SVM method performs worse than the other methods. The error floor of the proposed SVM-based channel estimator is lower than that of the BMMSE estimator, and the proposed SVM-based joint CE-DD method significantly improves the channel estimation accuracy. This is due to the help of the to-be-decoded data in refining the channel estimate.

In Fig.~\ref{fig_ChanEst_4K_32N_different_taus.eps}, we compare the NMSE of BMMSE with the NMSE of the proposed SVM-based method for different values of $T_\mathrm{t}$. It is observed that the high-SNR error floor of the BMMSE method quickly reaches a bound as $T_\mathrm{t}$ increases. However, the performance of the proposed SVM-based method improves as $T_\mathrm{t}$ increases. The error floor of BMMSE even with $T_\mathrm{t} = 100$ is still higher than that of the proposed SVM-based method with a much shorter training sequence ($T_\mathrm{t}=20$). The results in Fig.~\ref{fig_ChanEst_4K_32N_different_taus.eps} show that increasing $T_\mathrm{t}$ can help improve the channel estimation accuracy. However,  the spectral efficiency of the system is adversely affected as a result. Thus, the  proposed SVM-based joint CE-DD method can help improve both the channel estimation performance and the spectral efficiency. 

\begin{figure}[t!]
	\centering
	\includegraphics[width=\linewidth]{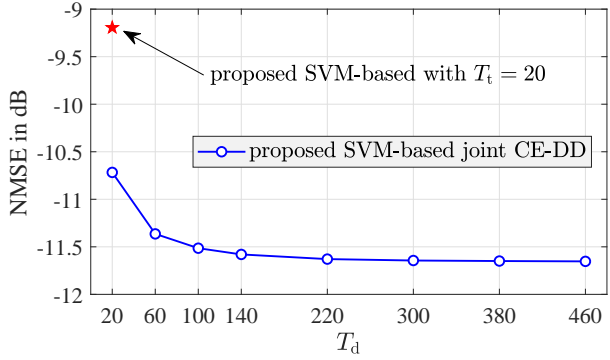}
	\caption{Effect of $T_\mathrm{d}$ on the $\mathrm{NMSE}$ of the proposed SVM-based joint CE-DD with $K = 4$, $N = 32$, and $T_\mathrm{t} = 20$ at $\rho = 30$ dB.}
	\label{fig_jointCEDD_4K_32N_different_Tb.eps}
\end{figure}

We study the effect of $T_\mathrm{d}$ on the NMSE of the proposed SVM-based joint CE-DD method in Fig.~\ref{fig_jointCEDD_4K_32N_different_Tb.eps}. It can be seen that as $T_\mathrm{d}$ increases, the channel estimation performance of the SVM-based joint CE-DD method reaches a bound. It is also seen that with a data segment of only about $150$ time slots, the channel estimation accuracy can asymptotically reach the bound, which is much better than the performance of using only the training sequence (the red star symbol).
\begin{figure}[t!]
	\centering
	\includegraphics[width=\linewidth]{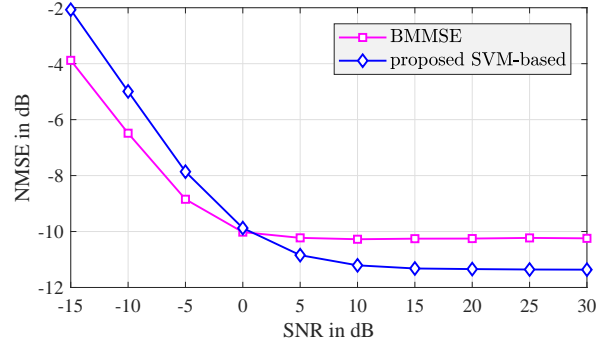}
	\caption{$\mathrm{NMSE}$ comparison between the BMMSE channel estimator and the proposed SVM-based channel estimator for spatially correlated channels with $K = 4$, $N = 32$, and $T_\mathrm{t} = 20$.}
	\label{fig_SpatiallyCorr_ChanEst_Comparison_4K_32N}
\end{figure}

Fig.~\ref{fig_SpatiallyCorr_ChanEst_Comparison_4K_32N} presents channel estimation results for spatially correlated channels. We use the same typical urban channel model as in~\cite{li2017channel}. The power angle spectrum of the channel model follows a Laplacian distribution with an angle spread of $10^{\circ} $. The simulation results indicate the performance advantage of the proposed SVM-based solution over the BMMSE method at high SNR, and thus justify the SVM-based problem formulation in \eqref{eq_SVM_correlated_chEst_formulation}.

\begin{figure}[t!]
	\centering
	\includegraphics[width=\linewidth]{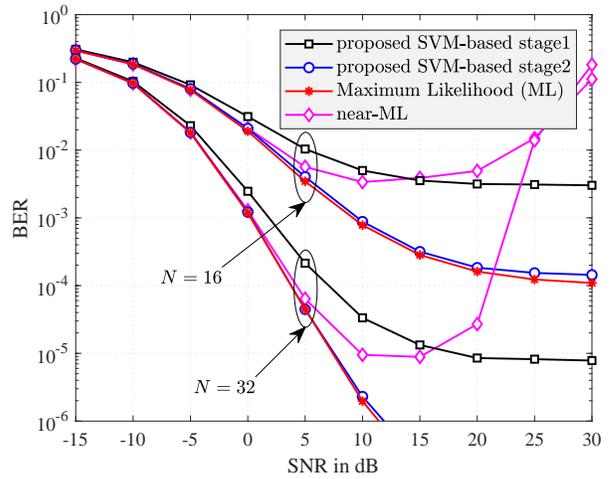}
	\caption{Performance comparison between the proposed two-stage SVM-based data detection method and ML detection \cite{choi2016near} with perfect CSI, QPSK modulation, and $K = 4$. The average cardinalities of $\mathcal{X}$ for $N = 16$ and $N = 32$ are $2.9352$ and $1.6140$, respectively.}
	\label{fig_BER_comparison_perfectCSI_4K}
\end{figure}
\begin{figure}[t!]
	\centering
	\includegraphics[width=\linewidth]{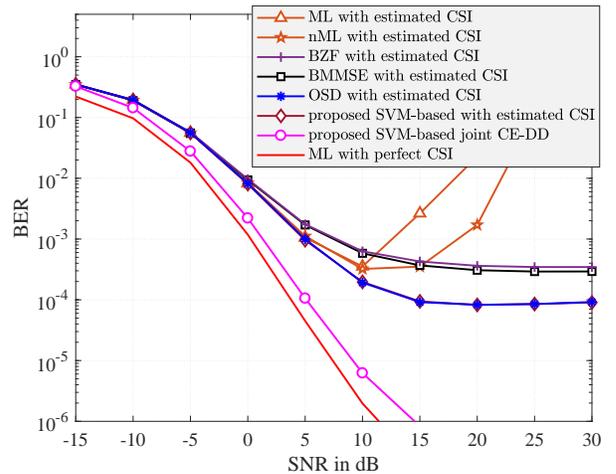}
	\caption{Performance comparison between two proposed data detection methods and other existing methods with estimated CSI, QPSK modulation, $N = 32$, $K = 4$, and $T_\mathrm{t} = 20$.}
	\label{fig_BER_comparison_imperfectCSI_32N_4K}
\end{figure}

In Fig.~\ref{fig_BER_comparison_perfectCSI_4K}, the proposed two-stage SVM-based data detection method is compared with the ML and nML detection methods for the case of perfect CSI. It is observed that the performance of the proposed method is very close to that of the ML method after two stages. It should be noted that the ML method performs well but it is an exhaustive-search method and so its computational complexity is prohibitively high for large-scale systems. While the nML method is applicable for large-scale systems, it is not robust at high SNRs. This non-robustness occurs regardless of the quality of the CSI, since nML depends on the gradient of a fractional form whose numerator and denominator both rapidly approach zero. It should also be noted that the average cardinalities of $\mathcal{X}$ for $N = 16$ and $N = 32$ are~$2.9352$ and~$1.6140$, respectively. This means the second stage of the proposed method is relatively simple to implement since it only has to search over a few candidates.

For the case of imperfect CSI, a bit-error-rate (BER) comparison is provided in Fig.~\ref{fig_BER_comparison_imperfectCSI_32N_4K}, where the estimated CSI is obtained by the SVM-based channel estimator. Here, the SVM-based joint CE-DD method can be compared with other methods because it also starts with CSI estimated by the SVM-based channel estimator. It is seen that both the ML and nML detection methods are non-robust at high SNRs with imperfect CSI. The susceptibility of ML was also reported in~\cite{Jeon2018supervised} and~\cite{nguyen2019supervised}. An explanation for the susceptibility of ML detection can be found in Appendix~\ref{sec_appendA}. It is also observed that the proposed SVM-based and OSD detection methods give the same performance. However, the proposed SVM-based joint CE-DD algorithm significantly outperforms other methods and its performance is quite close to the performance of the ML method with perfect CSI. This performance enhancement is due to the refined channel estimate obtained by solving~\eqref{eq_SVM_chEst_formulation2}.
\begin{figure}[t!]
	\centering
	\includegraphics[width=\linewidth]{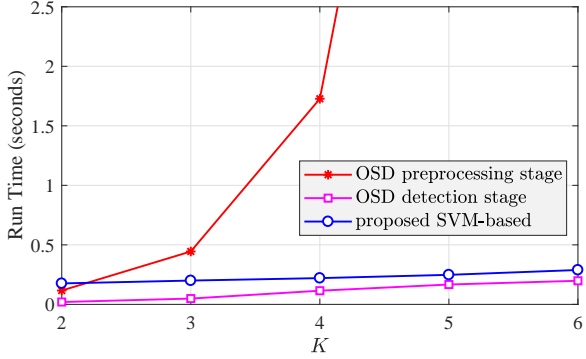}
	\caption{Run time comparison between OSD and the proposed SVM-based detection method with QPSK modulation, $N = 32$, and $K$ varies.}
	\label{fig_run_time_comparison}
\end{figure}

Although the SVM-based and OSD methods give the same performance, the computational complexity of the SVM-based approach is much lower than that of OSD. This is illustrated in Fig.~\ref{fig_run_time_comparison}. We calculate the average run time required to perform data detection over a block-fading interval of $500$ slots. Note that the OSD method contains two stages: a preprocessing stage and a detection stage. It is observed that the OSD method has a low-complexity detection stage. Interestingly, Fig. \ref{fig_run_time_comparison} indicates that the run time of proposed SVM-based method is comparable to that of the OSD detection stage. However, the OSD method requires a high-complexity preprocessing stage, which scales exponentially with the number of users. This makes the total complexity of the OSD method much higher than that of the SVM-based method, as observed in the figure.

\begin{figure}[t!]
	\centering
	\includegraphics[width=\linewidth]{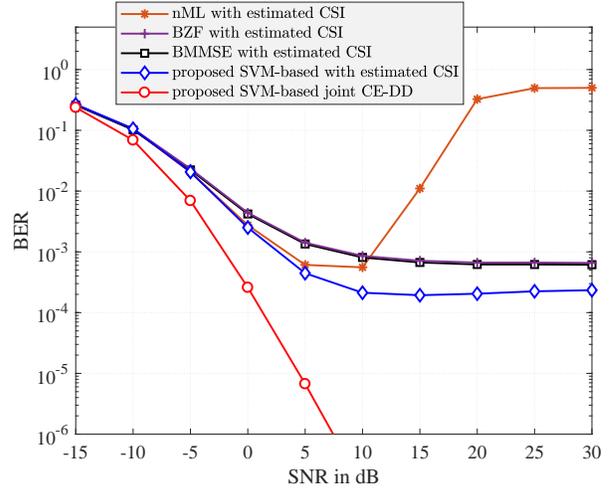}
	\caption{Performance comparison between two proposed data detection methods and other existing methods with estimated CSI, QPSK modulation, $N = 64$, $K = 8$, and $T_\mathrm{t} = 40$.}
	\label{fig_BER_comparison_imperfectCSI_QPSK_64N_8K}
\end{figure}

\begin{figure}[t!]
	\centering
	\includegraphics[width=\linewidth]{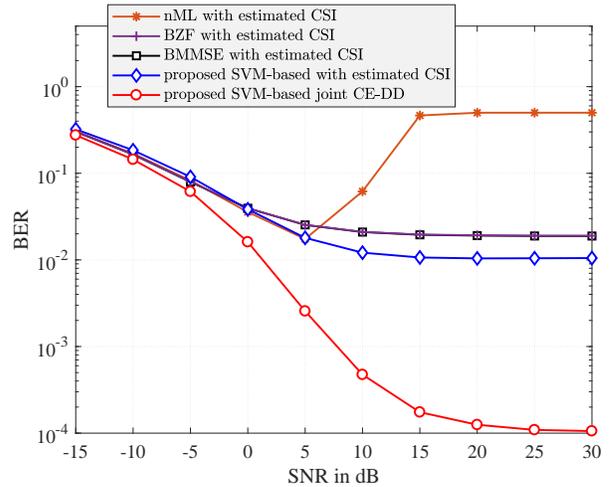}
	\caption{Performance comparison between two proposed data detection methods and other existing methods with estimated CSI, $16$-QAM modulation, $N = 128$, $K = 8$, and $T_\mathrm{t} = 40$.}
	\label{fig_BER_comparison_imperfectCSI_16QAM_128N_8K}
\end{figure}

Fig.~\ref{fig_BER_comparison_imperfectCSI_QPSK_64N_8K} and Fig.~\ref{fig_BER_comparison_imperfectCSI_16QAM_128N_8K} provide BER comparisons between the proposed SVM-based data detection methods and other existing methods with QPSK and $16$-QAM modulations using the CSI estimated by the SVM-based channel estimator. Due to their high computational complexity, we are not able to provide the BER of the ML and OSD detection methods. Instead, the performance of the nML method and other linear receivers are provided as alternatives. The proposed methods not only outperform the existing methods but are also robust at high SNRs.
\begin{figure}[t!]
	\centering
	\includegraphics[width=\linewidth]{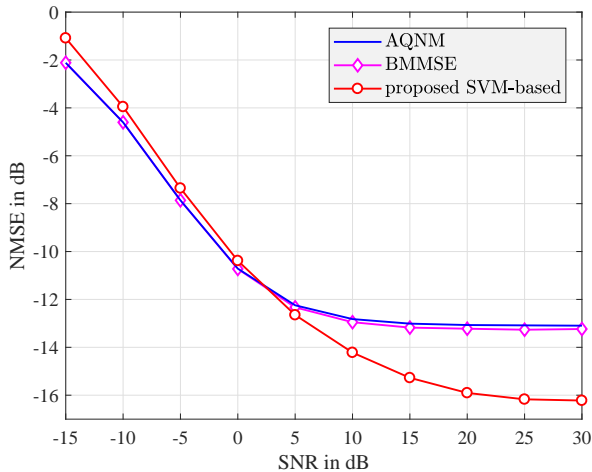}
	\caption{$\mathrm{NMSE}$ comparison between different channel estimators for an OFDM system in a frequency-selective channel with $K = 2$, $N = 16$, and $L=8$.}
	\label{fig_FreqSelec_ChanEst_Comparison_2K_16N_8L_256Nc}
\end{figure}

\begin{figure}[t!]
	\centering
	\includegraphics[width=\linewidth]{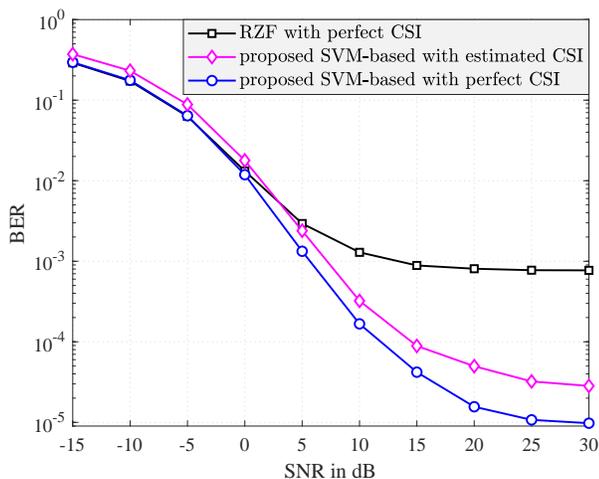}
	\caption{$\mathrm{BER}$ comparison between different data detection methods for an OFDM system in a frequency-selective channel with $N_c=256$, QPSK modulation, $K = 2$, $N = 16$, and $L=8$.}
	\label{fig_FreqSelec_DataDetect_QPSK_2K_16N_8L_256Nc}
\end{figure}

Finally, channel estimation and data detection results for OFDM systems with frequency-selective fading channels are given in Fig.~\ref{fig_FreqSelec_ChanEst_Comparison_2K_16N_8L_256Nc} and Fig.~\ref{fig_FreqSelec_DataDetect_QPSK_2K_16N_8L_256Nc}, respectively. It is observed that the BMMSE channel estimator~\cite{li2017channel} slightly outperforms the AQNM-based channel estimator~\cite{Mollen2017Uplink}, but both of these methods have higher NMSE than the proposed SVM-based channel estimator at high SNRs. More specifically, the high-SNR error floor of the SVM-based method is about $3$-dB lower that that of the BMMSE and the AQNM-based methods. In Fig.~\ref{fig_FreqSelec_DataDetect_QPSK_2K_16N_8L_256Nc}, data detection results show that the SVM-based method considerably outperforms the Regularized Zero-Forcing (RZF) of~\cite{Mollen2017Uplink}. At high SNRs, the BER of the RZF method even with perfect CSI is much higher than the BER of the SVM-based method with estimated CSI.
\section{Conclusion}
\label{sec_conclusion}
In this paper, we have shown how linear SVM, a well-known machine learning technique, can be exploited to provide efficient and robust channel estimation and data detection. We proposed SVM-based channel estimation methods for both uncorrelated and spatially correlated channels, a two-stage SVM-based data detection method, and an SVM-based joint CE-DD method. Extension of the proposed methods to OFDM systems with frequency-selective fading channels was also derived. The key idea is to formulate the channel estimation and data detection problems as SVM problems so that they can be efficiently solved. Simulation results revealed the superiority of the proposed methods against existing ones and the gain is greatest for moderate to high SNR regimes.

\appendices
\section{Explanation for the susceptibility of ML detection at high SNRs with imperfect CSI}
\label{sec_appendA}
The ML detection method of~\cite{choi2016near} is defined as
\begin{equation}
\hat{\mathbf{x}}^{\mathtt{ML}}_{\mathrm{d},m} = \argmax_{\bar{\mathbf{x}}\in \mathcal{M}^{N_\mathrm{t}}} \underbrace{\prod_{i=1}^{2N} \Phi\left(\sqrt{2\rho} y_{\mathrm{d},m,i}\hat{\mathbf{h}}_{\mathrm{d},i}^T\mathbf{x}\right)}_{\mathcal{L}(\mathbf{x})},
\label{eq_ML_detection}
\end{equation}
where $\mathbf{x} = [\Re \{\bar{\mathbf{x}}\}^T, \Im \{\bar{\mathbf{x}}\}^T]^T$, $\mathcal{L}(\mathbf{x})$ is the likelihood function, and $\Phi(t) = \int_{-\infty}^{t}\frac{1}{\sqrt{2\pi}}e^{-\tau^2/2}d\tau$ is the cumulative distribution function of the standard Gaussian random variable. It is clear that as $\rho\rightarrow\infty$, we have
\begin{equation*}
\begin{cases}
\Phi\left(\sqrt{2\rho} y_{\mathrm{d},m,i}\hat{\mathbf{h}}^T_{\mathrm{d},i}\mathbf{x}\right) \rightarrow 0 \text{ if } y_{\mathrm{d},m,i}\hat{\mathbf{h}}^T_{\mathrm{d},i}\mathbf{x} < 0,\\
\Phi\left(\sqrt{2\rho} y_{\mathrm{d},m,i}\hat{\mathbf{h}}^T_{\mathrm{d},i}\mathbf{x}\right) \rightarrow 1 \text{ if } y_{\mathrm{d},m,i}\hat{\mathbf{h}}^T_{\mathrm{d},i}\mathbf{x} > 0.
\end{cases}
\end{equation*}
This means, as $\rho \rightarrow \infty$, $\mathcal{L}(\mathbf{x}) = 0$ if there exists at least one index $i$ such that $y_{\mathrm{d},m,i}\hat{\mathbf{h}}^T_{\mathrm{d},i}\mathbf{x} < 0$ and $\mathcal{L}(\mathbf{x}) = 1$ if  $y_{\mathrm{d},m,i}\hat{\mathbf{h}}^T_{\mathrm{d},i}\mathbf{x} > 0$ for all $i$.

Now, suppose that a vector $\bar{\mathbf{x}}^\star$ was transmitted and let $\mathbf{x}^\star = [\Re \{\bar{\mathbf{x}}^\star\}^T, \Im \{\bar{\mathbf{x}}^\star\}^T]^T$. If the CSI is perfectly known, i.e., $\hat{\mathbf{h}}_{\mathrm{d},i} = \mathbf{h}_{\mathrm{d},i}$, we have $y_{\mathrm{d},m,i}\hat{\mathbf{h}}^T_{\mathrm{d},i}\mathbf{x}^\star > 0$ for all $i$ because $y_{\mathrm{d},m,i} =  \operatorname{sign}(\mathbf{h}^T_{\mathrm{d},i}\mathbf{x}^\star) = \operatorname{sign}(\hat{\mathbf{h}}^T_{\mathrm{d},i}\mathbf{x}^\star)$ as $\rho \rightarrow \infty$. In other words, $\mathcal{L}(\mathbf{x}^\star) = 1$ if the CSI is perfectly known at infinite SNR. However, if the CSI is not known perfectly, i.e., $\hat{\mathbf{h}}_{\mathrm{d},i} \neq \mathbf{h}_{\mathrm{d},i}$, there is a non-zero probability that $y_{\mathrm{d},m,i}=  \operatorname{sign}(\mathbf{h}^T_{\mathrm{d},i}\mathbf{x}^\star)\neq \operatorname{sign}(\hat{\mathbf{h}}^T_{\mathrm{d},i}\mathbf{x}^\star)$, which means $y_{\mathrm{d},m,i}\operatorname{sign}(\hat{\mathbf{h}}^T_{\mathrm{d},i}\mathbf{x}^\star) < 0$. This causes $\mathcal{L}(\mathbf{x}^\star) = 0$. For any $\mathbf{x} \neq \mathbf{x}^\star$, it is possible that $y_{\mathrm{d},m,i}=  \operatorname{sign}(\mathbf{h}^T_{\mathrm{d},i}\mathbf{x}^\star)\neq \operatorname{sign}(\hat{\mathbf{h}}^T_{\mathrm{d},i}\mathbf{x})$, which also leads to $\mathcal{L}(\mathbf{x}) = 0$. Hence, detection errors occur. The above explanation is argued at infinite SNR, but it is also valid for high SNRs because the function $\Phi(t)$ approaches~$0$ very fast.

To remove the product in~\eqref{eq_ML_detection}, one may argue to transform the function $\mathcal{L}(\mathbf{x})$ into a sum of $\log$ functions as follows:
\begin{equation}
\hat{\mathbf{x}}^{\mathtt{ML}}_{\mathrm{d},m} = \argmax_{\bar{\mathbf{x}}\in \mathcal{M}^{N_\mathrm{t}}} \underbrace{\sum_{i=1}^{2N} \log \Phi\left(\sqrt{2\rho} y_{\mathrm{d},m,i}\hat{\mathbf{h}}_{\mathrm{d},i}^T\mathbf{x}\right)}_{\mathcal{L}(\mathbf{x})}.
\label{eq_MlogL_detection}
\end{equation}
However, the function $\mathcal{L}(\mathbf{x})$ in~\eqref{eq_MlogL_detection} still depends on $\Phi(\cdot)$ and can involve $\log(0)$. The proposed SVM-based data detection method is robust against imperfect CSI since it does not depend on the $\Phi(\cdot)$ function and information about the SNR is not required either.

We note that the OSD method in~\cite{Jeon2018One} is also robust against imperfect CSI thanks to the use of the approximation $\Phi(t) \approx \frac{1}{2}e^{-0.374t^2-0.777t}$ for non-negative $t$. This approximation helps remove the effect of $\log \Phi(\cdot)$ in~\eqref{eq_MlogL_detection} since $\log e^a = a$. However, the OSD method has higher computational complexity than the proposed SVM-based methods.
\bibliographystyle{IEEEtran}
\bibliography{ref}

\end{document}